# Enabling high-precision 3D strong-field measurements - Ionization with low-frequency fields in the tunneling regime

J. Dura[1], N. Camus[2], A. Thai[1], A. Britz[1], M. Hemmer[1], M. Baudisch[1], A. Senftleben[2], C.D. Schröter[2], J. Ullrich[2,3], R. Moshammer[2], J. Biegert[1,4*]

[1]ICFO-Institut de Ciences Fotoniques, 08860 Castelldefels (Barcelona), Spain

[2]Max-Planck-Institut für Kernphysik, Saupfercheckweg 1, 69117 Heidelberg, Germany

[3]Physikalisch-Technische Bundesanstalt, D-38116 Braunschweig, Germany

[4]ICREA – Institució Catalana de Recerca i Estudis Avançats, 08010 Barcelona, Spain

*Correspondence to: jens.biegert@icfo.eu

**Ionization of an atom or molecule presents surprising richness beyond our current understanding: strong-field ionization with low-frequency fields recently revealed unexpected kinetic energy structures (1, 2). A solid grasp on electron dynamics is however pre-requisite for attosecond-resolution recollision imaging (3), orbital tomography (4), for coherent sources of keV light (5), or to produce zeptosecond-duration x-rays (6). We present a methodology that enables scrutinizing strong-field dynamics at an unprecedented level. Our method provides high-precision measurements only 1 meV above the threshold despite 5 orders higher ponderomotive energies. Such feat was realized with a specifically developed ultrafast mid-IR light source in combination with a reaction microscope. We observe electron dynamics in the tunneling regime ($\gamma$ = 0.3) and show first 3D momentum distributions demonstrating surprising new**

**observations of near-zero momentum electrons and low momentum structures, below the eV, despite quiver energies of 95 eV.**

Full scrutiny of strong field processes requires a light source reaching relevant field strengths with repeatable electric field waveforms, and 3D momentum detection. Mid-IR probing removes ambiguities between tunneling and multi-photon processes, but unfavorable scaling of cross-sections, and lack of mid-IR ultrafast technology, has largely prevented this approach. We now enable such measurements with a specifically designed mid-IR light source (*7*; see Methods), which fulfills all requirements and is perfectly matched to a reaction microscope thereby permitting rapid and high-precision data acquisition. We show here the power of our method by applying it to the recent unexpected low-energy structures (LES).

The LES was first observed by Blaga et al. (*1*) at several eV, when ionizing argon and xenon atoms (also molecular hydrogen and nitrogen) with short pulses at long wavelengths. Quan et al. (*2*) showed similar results shortly thereafter for xenon. Recently, Wu et al. (*8*) observed even slower electrons (just below 1 eV), for similar conditions in krypton and xenon, termed very low energy structure (VLES). The VLES seems to persist for 800 nm in neon even though it is hard to identify. It is difficult to disentangle the origin of the various findings without momentum distribution measurements for LES or VLES; moreover, experimental conditions, varied across these investigations. Without momentum information, the appearance of VLES, and for LES, was attributed to long-range Coulomb effects.

On the theory side, models based on the strong-field approximation (SFA) (*1*) were unable to predict the LES. Comparison of SFA, which neglects the ion's Coulomb field, with ab-initio calculations supported the assumption of the ion's Coulomb field as responsible for the modification of electron trajectories (*1, 9-14*). Direct tunneling close to a local field maximum, even though resulting in low-energy photoelectrons (until $2U_p$; ponderomotive energy $U_p = E^2/4\omega^2$; $E$ is the peak electric field amplitude and $\omega$ the laser frequency in atomic units), is therefore seen as an unlikely cause of the LES. The location of the LES, at energies far below $2U_p$, would exclude rescattering ($2U_p$ to $10U_p$) as underlying mechanism, while its disappearance with circularly polarized light (*1*) would contradict this notion. Faisal pointed out (*15*) that forward scattering from the successive half-cycle, could be another competing channel, which should be considered. Several theoretical investigations quickly confirmed the strong contribution from forward-scattering (*9, 10, 12, 14*) and, therefore, identified the LES as resulting from soft recollisions modified by the interaction of the outgoing electron with the ion's Coulomb field (*9, 10*). Interestingly, even though momentum distributions were calculated, no investigation was concerned with extremely slow electrons (VLES) and the discussion has not been settled yet (*16*).

To add to the confusion, in a recent publication Liu et al. (*17*) report opposing experimental findings: they observe low yield of low kinetic energy electrons from strong-field ionization of krypton and xenon atoms using 1320 nm light and at similar pulse peak intensity as in the previously cited experimental papers. The authors attribute their contradicting observations to population trapping in high-lying (Rydberg) states as suppression mechanism. In fact, the possibility of an atom resisting ionization by a strong laser field has been pointed out early on by Fedorov (*18*) and others (*19*) in the multi-photon picture and in the tunneling regime (20, 21); this mechanism was termed as frustrated tunnel ionization (*22*). We note that, aside from the contradicting measurements, the existence of such local ionization suppression is under debate and some authors question the mechanism altogether (*23*).

Thus, one is faced with a situation that there are surprising, new, and inconsistent findings on low-energy photoelectron spectra and angular distributions in the mid-IR laser wavelength regime that were neither anticipated by theory nor are they consistently explained. Thus, the goal of our investigation was to clarify this situation with unambiguous and comprehensive high-resolution 3D electron momentum data for strong-field ionization by low-frequency laser fields for argon atoms and oxygen molecules. Ar and $O_2$, resemble the most thoroughly scrutinized atomic and simple molecular systems, therefore providing a benchmark for future theoretical and experimental discussions. We specifically elect $O_2$, and not argon's molecular partner $N_2$, since nitrogen is known to exhibit similar response to argon upon ionization, while $O_2$ exhibits a very different response due to its different electronic structure. Our measurements have become possible in the non-perturbative tunneling regime (Keldysh adiabaticity parameter (*25*) $\gamma = \sqrt{I_p/2U_p} = 0.3 \ll 1$, $U_p$ = 95 eV) due to the development of a unique high-repetition rate (160 kHz) source of intense and carrier-to-envelope phase (CEP) stable, few-cycle pulses in the mid-IR (*24*) (6 cycle duration, 3100 nm (0.4 eV or 0.015 a.u), 16 μJ), which is pre-requisite for 3D momentum imaging with a reaction microscope (REMI); details on the source and detector can be found in Methods.

Our kinematically complete measurement permits an unprecedented view at low-frequency photoionization enabling identification of its underlying contributions. We find striking new features in Fig. 1(a) for Ar ($P_{parallel}$ < 0.4 a.u. (2.18 eV), and $P_{transverse}$

< 0.5 a.u. (3.4 eV); similarly for oxygen in Suppl. Fig.1) such as an extremely narrow contribution ("zero-peak") of zero, or near-zero momentum electrons (I) and, emanating from there, a "v-like" pattern (II) which widens and merges with the broad distribution (III) at larger transverse momenta. We analyze the kinetic energy spectrum and parallel momentum distribution for the individual areas (I, II, III) and identify the broad contribution of area III (0.06 a.u. to 0.3 a.u transverse momentum) as LES according to the kinetic energy spectrum (Fig. 1(e)). Associated with this area is a broad double hump structure in parallel momentum (Fig. 1(d)) that is significantly narrower (Fig. 1(f)) for area II. The kinetic energy spectrum (Fig. (g)) permits identifying the "v-like" distribution as the VLES. Most striking is the convergence of the "v-like" structure into an extremely narrow contribution extending towards zero momentum (I) despite quiver energy of 95 eV. The offset of the "zero-peak" from zero transverse momentum is within our measurement resolution of (0.008 ± 0.026) a.u. in the transverse and |(0 ± 0.012)| a.u. in the parallel direction.

Our measurement allows, for the first time, disentangling parallel momentum contributions as originating from VLES and LES. The persistence of VLES at shorter wavelength (*8*) and the absence of LES would indicate that the previously found double-hump structures (*26, 27*) are reminiscent of VLES. Areas (I) and (II) show never-observed features which we find as being of very general nature for the different target species investigated and varying in their relative contributions. Measurements in oxygen, Suppl. Fig. 1, show less slow electrons, but the zero kinetic-energy feature is clearly visible. Our observation is consistent with previous findings of Rudenko et al. (*26*) and Liu et al. (*27*) who have observed a zero momentum peak in the parallel ion momentum of argon and oxygen respectively but not for other molecules and noble gases.

Both features (Fig. 1(f,g) and (h,i)) disappear when introducing 15% ellipticity (see Fig. 2). Figure 2 shows measurements of the 3D integrated photoelectron kinetic-energy spectrum for linear and elliptically polarized light. We find, analogously to Ref. (*1*), the LES peaks at higher kinetic energy for elliptically polarized light and our newly observed feature at 1.3 meV is not discernible any more. Note that the peak at 1.3 meV is within our experimental resolution of 11 meV. Persistence of the "zero-peak" across various species, and its disappearance when ionizing with circularly polarized light suggests involvement of electron rescattering (*9-16*). What could the cause for the low momentum be? Trapping of population in high-lying, or Rydberg, states (*13,20,21*) could, ultimately, result in electrons (*19*) with very small momenta (*28*). Figure 3 indeed evidences strong interaction with the ion's field, which could be responsible for electron trapping after tunneling (*20,22*): deviation from a Gaussian-like distribution, in favor of a cusp-like distribution, in the perpendicular electron momenta is the unambiguous signature of the Coulomb field (*29*). We find the lowest energy electrons show the strongest deviation (Fig. 1(h)); analog findings are shown in Suppl. Fig. 2 for $O_2$ molecules. The hypothesis of involvement of high-lying states, ultimately leading to low-momentum electrons, is also supported by elliptically polarized light reducing population in high-lying states (*22, 30*); this would explain the absence of low kinetic-energy electrons with elliptically polarized light.

Our method enables a new level of high precision strong field measurements and sheds new light on recollision dynamics with striking new findings. We observe unseen "v-like" patterns in the electron's momentum distribution and exceptionally slow electrons at near-zero momentum. These measurements reveal a much more intricate and less trivial behavior of atoms and simple molecules under strong mid-IR fields than previously observed or expected. The persistence of these findings is independent on electronic structure and hints at quantum interference, due to the strong external field with the ion's Coulomb field, and the importance of transient intermediate states. We are currently working in this direction to further corroborate our findings. Our results are important for realizing the enticing possibilities of mid-IR rescattering physics, which require a firm grasp on the underlying electron dynamics.

**Methods:**

**Mid-IR Light Source**.

The experiments were performed with our home-built mid-IR optical parametric chirped pulse amplifier (OPCPA) (*7*), which delivers linearly polarized, 60-fs (6-cycle) pulses at 3100 nm centre wavelength and 160 kHz pulse repetition rate. Pulse energies of up to 16 µJ are achieved after compression with a power stability of better than 0.3% rms and 0.8% peak-to-peak over 15 hours. The OPCPA is optically carrier-to-envelope phase (CEP) self-stabilized and routinely achieves 250 mrad stability over 30 min. We refer for a detailed description of our OPCPA light source to Refs. (*24, 7*).

**Reaction Microscope**.

High-pressure gas is supersonically expanded into vacuum and skimmed in two successive stages to deliver a cold and directed atomic beam into the target chamber of the REMI. We use a mass flow controller to ensure identical number density of $10^{11}/cm^3$ for all target species used in our experiments. The target momentum resolution is estimated as 0.03 a.u. in jet direction and 0.012 a.u. perpendicularly. Momentum accuracy is 0.025 a.u. transversally and 0.012 a.u. parallel to the laser field. 8 µJ-energy pulses are focused with an on-axis, gold-coated, 1.2 f-number paraboloid into the gas jet, forming a 6 µm-width focal region. Upon ionization, electrons and ions are separated by a weak electric field (1.5 V/cm) and guided towards opposing micro channel plate (MCP) detectors equipped with delay line readout. Helmholtz coils generate a homogenous magnetic field (460 µT) to ensure near 4π collection efficiency for the fragments of interest. The jet is extracted separately to ensure a backing pressure of the REMI of $6 \times 10^{-9}$ mbar (without jet $3 \times 10^{-11}$ mbar). Time-of-flight (TOF) information is extracted from the MCP detectors, which are triggered from a fast photodiode (PD). The delay-line readout provides position information and the combination with TOF permits reconstructing 3D momentum vectors for ions and electrons in coincidence. We operate the gas jet at a density low enough to ensure an average count rate of 0.03 per pulse in order to minimize false coincidences. Data was collected over duration of 21 hours at the full repetition rate of the laser system at 160 kHz and in coincidence. The data set for argon contains $9 \times 10^6$ coincidence events.

**Acknowledgments** We acknowledge support from the Spanish Ministerio De Economia Y Competitividad (MINECO) through its Consolider Program (SAUUL-CSD 2007-00013), "Plan Nacional" (FIS2011-30465-C02-01) and the Catalan Agencia de Gestió d'Ajuts Universitaris i de Recerça (AGAUR) with SGR 2009-2013. This research has been partially supported by Fundació Cellex Barcelona, LASERLAB-EUROPE, grant agreement 228334 and COST Action MP1203. J.D. was partially supported by FONCICYT Project 94142.

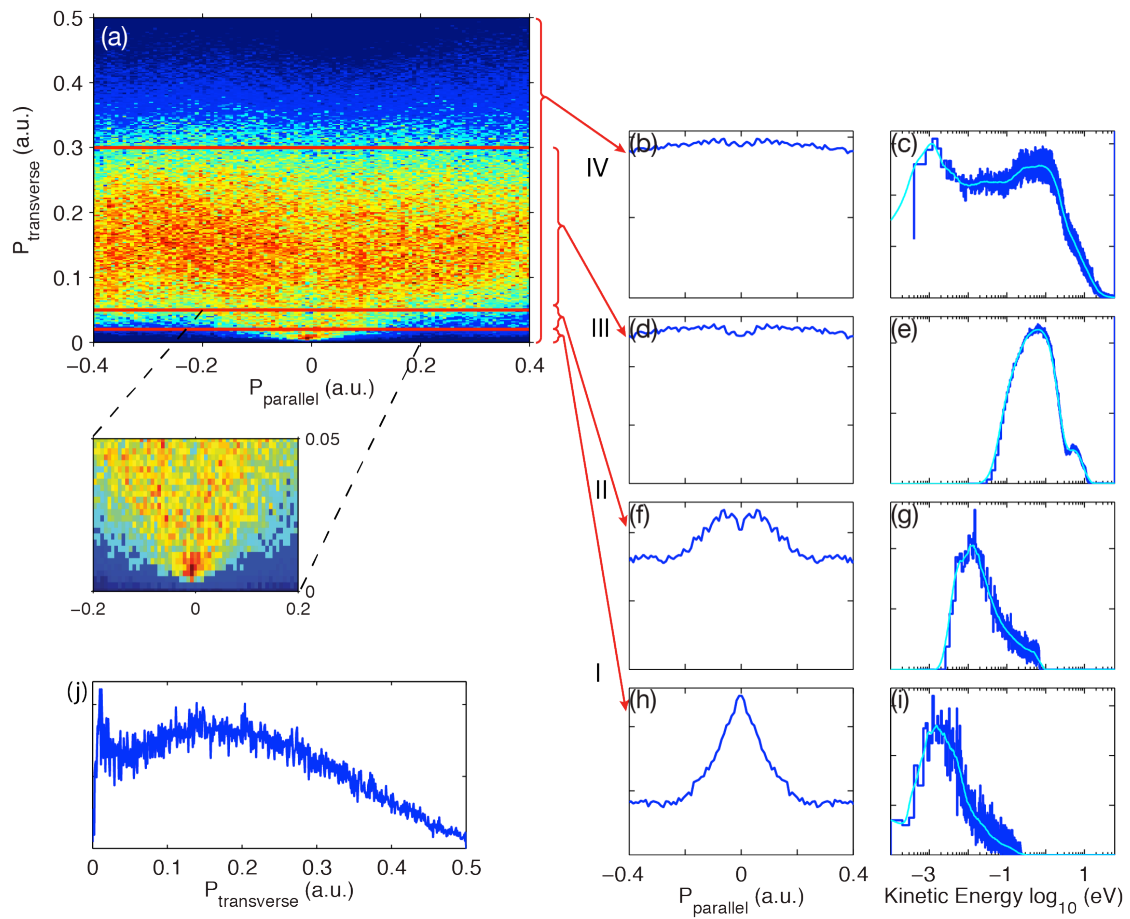

**Figure 1 | Electron momentum distributions and correlated total kinetic energy for Ar**. Graph (b) shows the parallel electron momentum and (c) the angle integrated electron kinetic energy for the complete distribution (a). The following graphs highlight contributions from the various areas (I, II and III) within the momentum map (a) to parallel electron momenta (b,d,f,h) and angle integrated kinetic energy (c,e,g,i). Zero-, or near-zero-momentum electrons (I) contribute in the meV to the kinetic energy spectrum (i), to a peak at zero parallel electron momentum (h) and to a peak at zero-, or near-zero-transverse momentum (j). A "v"-like structure (II) is visible at slightly larger transverse momenta, corresponding to a narrow double-hump structure in the parallel electron momentum spectrum (f) which is different from the

broad momentum distribution (e), associated with LES, which results in a broad hump-like structure (d).

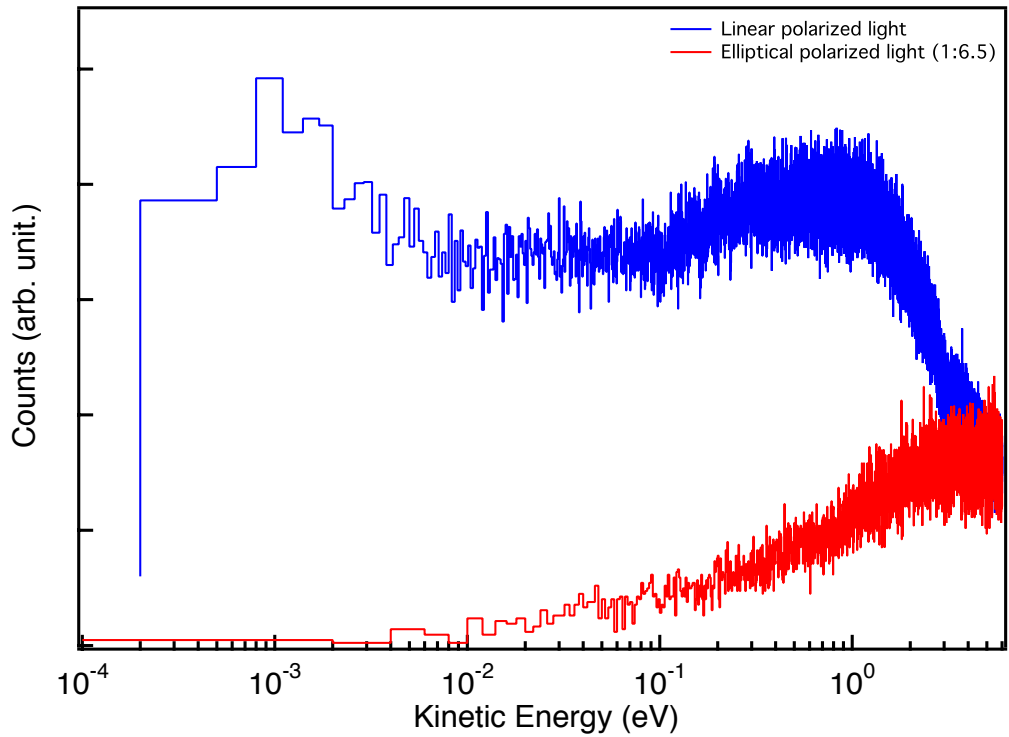

**Figure 2 | Photoelectron kinetic energy for Ar**. Total kinetic energy spectrum for the photoelectrons, integrated over 4π steradian. Shown is the case for irradiation with linear polarization and 15% ellipticity; both for randomized CEP.

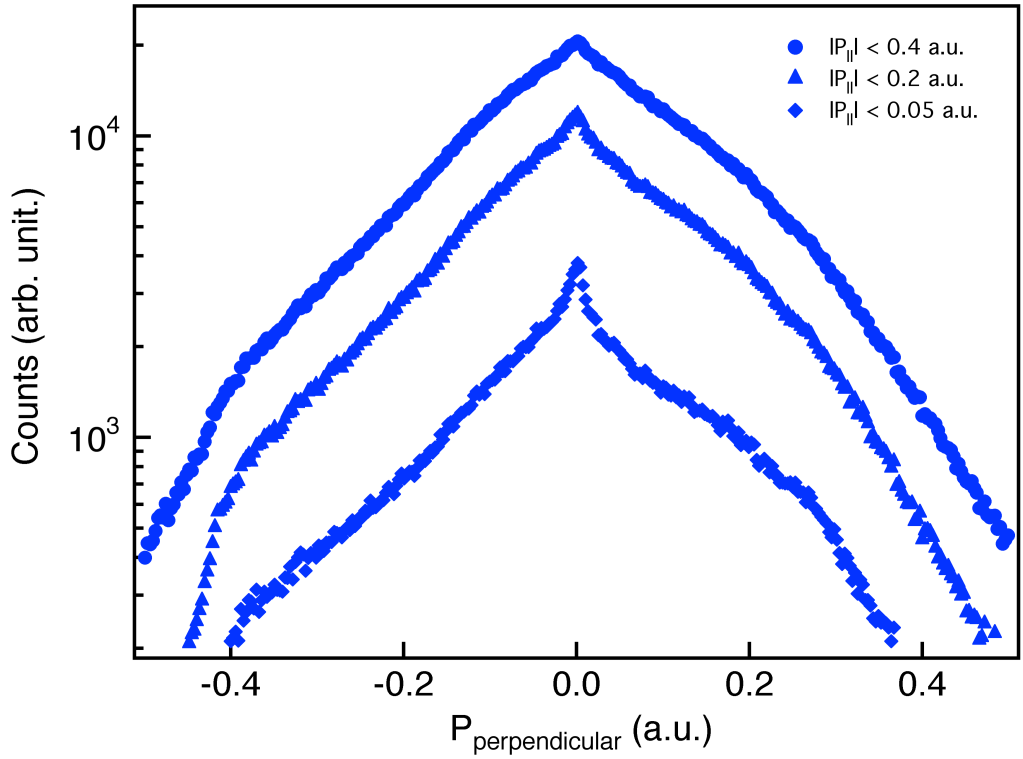

**Figure 3 | Perpendicular electron momentum distributions for Ar.** One component of the electron momentum, perpendicular to the laser polarization, for different integration ranges along the parallel momentum as indicated in the figure and highlighted by different symbols. The strong influence of the Coulomb potential manifests itself for the smallest longitudinal momentum cut in form of a strong cusp-like feature (diamonds).

# Supplementary Information

To manuscript titled:

**"Strong field ionization with low-frequency fields in the tunneling regime"**

By authors: J. Dura, N. Camus, A. Thai, A. Britz, M. Hemmer, M. Baudisch, A. Senftleben, C.D. Schröter  J. Ullrich, R. Moshammer, J. Biegert.

**Oxygen molecule results**

Here we show measurements for identical laser parameters but for randomly oriented oxygen molecules. The different ionization potential results in slightly different parameters $\gamma = 0.25$, $z = 237$ and $z_1 = 15.7$. The measurement in Supplementary Fig. 1 shows $5 \times 10^6$ coincidence events and similar qualitative behavior as for argon.

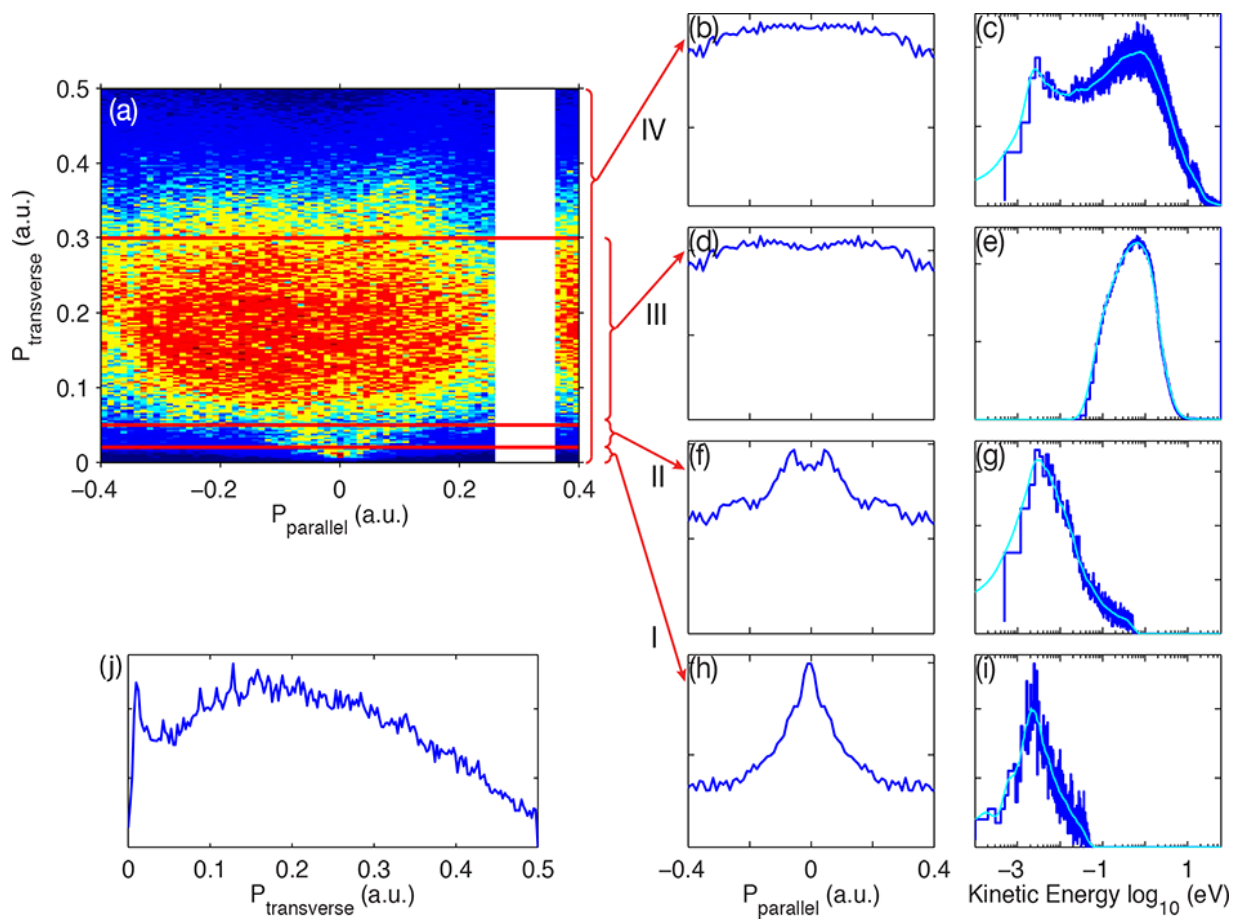

**Supplementary Fig. 1 | Electron momentum distributions and correlated total kinetic energy for $O_2$.** Electron momentum distributions (a), parallel electron momenta (c,d,f,h) and correlated total kinetic energy for $O_2$ (c,e,g,i) as well as transverse electron momentum (j).

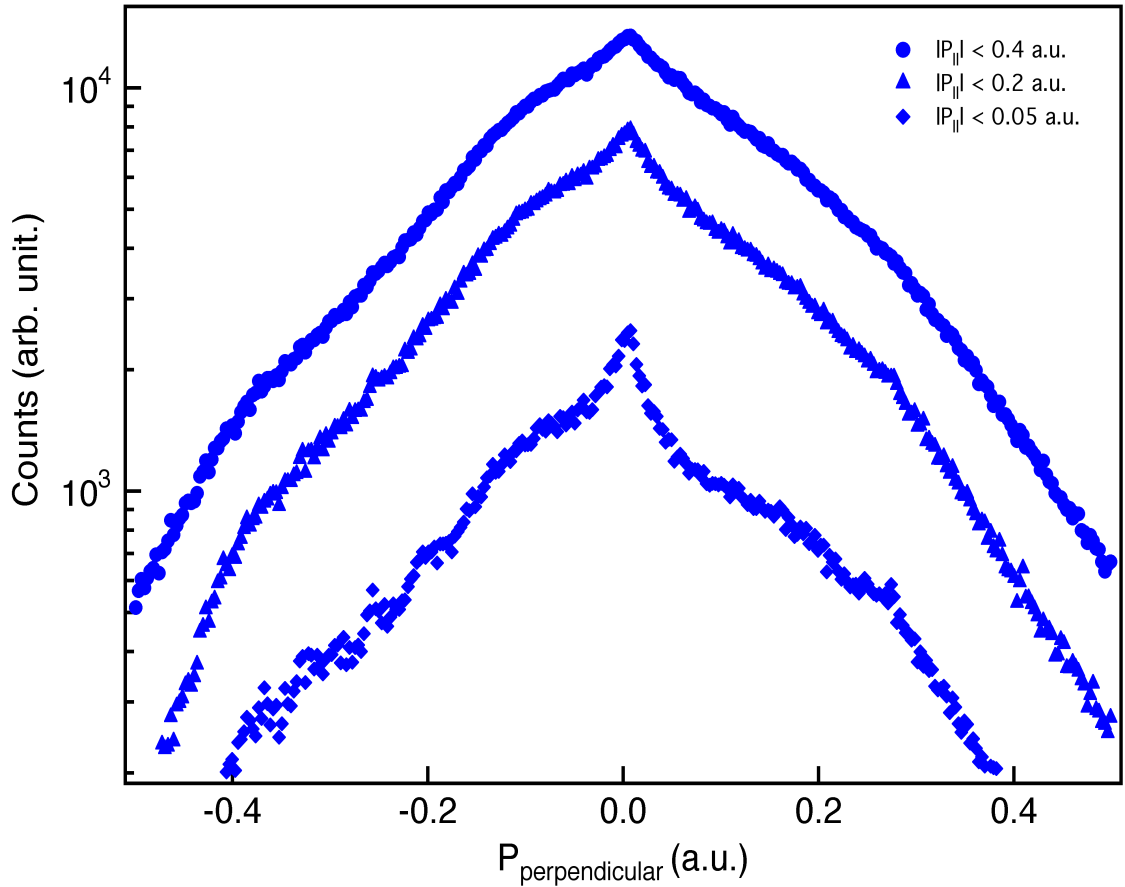

**Supplementary Fig. 2 | Perpendicular electron momentum distributions for $O_2$.** One component of the electron momentum, perpendicular to the laser polarization, for different integration ranges along the parallel momentum as indicated in the figure and highlighted by different symbols. The strong influence of the Coulomb potential manifests itself for the smallest longitudinal momentum cut in form of a cusp-like feature (diamonds).